\documentstyle[epsfig,psfrag]{aipproc}

\newcommand{\eq}[1]{equation~(\ref{#1})}

\newcommand{\be}{\begin{equation}}
\newcommand{\ee}{\end{equation}}

\def\alphaT {\alpha_{,\eta}}
\def\betaT {\beta_{,\eta}}
\def\PT {\phi_{,\eta}}

\def\alphaZ {\alpha_{,\zeta}}
\def\betaZ {\beta_{,\zeta}}
\def\PZ {\phi_{,\zeta}}

\begin{document}
\title{Inhomogeneity and the Post-Inflationary Universe}

\author{Richard Easther and Matthew Parry}
\address{Department of Physics, Brown University, Providence, 
RI02912}

\maketitle

\begin{abstract}
We discuss the interaction between perturbations in the inflaton and
the background during the preheating phase in simple inflationary
models. By numerically solving the Einstein field equations we are
able to assess the impact of non-linear gravitational effects on
preheating, and to assess the accuracy of perturbative discussions of
the preheating epoch.  
\end{abstract} 

\section*{Introduction}

Inflation drives the primordial universe towards the very special
initial state needed for it evolve into the form it is observed to
have today.  However, to be successful, inflation must terminate
gracefully with the energy density of the inflaton field being
transferred to radiation, a process known as {\em reheating\/}.

Since 1990 \cite{TraschenET1990a} it has been realized that reheating
can be driven by nonlinear, coherent effects leading to explosive
particle production via parametric resonance. The resulting
distribution is far from thermal equilibrium, and the process is often
called {\em preheating\/} (see \cite{KofmanET1997a,ParryET1998a}, and
refs. within).

The standard analytic approach to preheating is based on a Fourier
expansion of the relevant fields. The equation of motion of the $k$-th
mode has the form of a Mathieu or Lam\'{e} equation. Modes
corresponding to values of $k$ in resonance bands grow exponentially,
and the precise resonance structure depends sensitively on the
underlying particle physics.  Because it selectively amplifies
specific Fourier modes, parametric resonance implies an enhancement in
the spatial variation of the fields.  While the non-linear
backreaction on the field evolution has been examined analytically
\cite{KofmanET1997a,GreeneET1997a} and numerically
\cite{KhlebnikovET1996a,ProkopecET1996a}
the backreaction of the inhomogeneity on the underlying spacetime
metric and its implications for the evolution of the universe has only
recently begun to be studied
\cite{ParryET1998a,%
KodamaET1996a,%
HamazakiET1996a,%
NambuET1996a,%
BassettET1998a,%
FinelliET1998a,%
BassettET1999a,%
EastherET1999a}.

We have broken new ground by tackling the back-reaction problem using
the full Einstein field equations, and eschewing the use of any
perturbative approximations.  We show that large metric
inhomogeneities can be induced by parametric resonance, so
approximations to the full field equations may not adequately describe
the evolution of the universe during (and after) preheating.  At
present, our principal simplifying assumption is that the
inhomogeneity lies in a single spatial direction, which allows us to
work with a $1+1$ dimensional system of partial differential
equations.

Here we briefly survey the growth of metric perturbations during
preheating in an inflationary model driven by a $\lambda \phi^4$
potential. When treated analytically, the $m^2\phi^2$ model does not
undergo parametric resonance during reheating \cite{FinelliET1998a}, a
result we confirmed numerically \cite{ParryET1998a}. However, $\lambda
\phi^4$ possesses a single resonance band, and when we examine the
evolution of modes within this band we see observe strong
amplification of the corresponding metric perturbations.  This work
will be more fully described in our forthcoming paper
\cite{EastherET1999a}.

\section*{Metric and Initial Conditions}

We have assumed that the universe has a planar symmetry, so the metric
functions depend only on $t$ and $z$, and are independent of $x$ and
$y$. In \cite{ParryET1998a}, we studied reheating after $m^2 \phi^2$
inflation using the metric
\be \label{metric}
ds^2 = dt^2 - A^2(t,z)\,dz^2 - B^2(t,z)\,(dx^2 + dy^2), 
\ee
which describes an inhomogeneous universe in which the $dx^2 + dy^2$
sections have zero spatial curvature.  In principle, we could retain
this metric for the $\lambda \phi^4$ case, but it is advantageous to
work in conformal-like co-ordinates analogous to those which simplify
the homogeneous system \cite{GreeneET1997a,EastherET1999a}.
Specifically, by transforming $t$ and $z$ to $\eta$ and $\zeta$ we
write \eq{metric} as:
\be\label{metric2}
ds^2 = \alpha^2(\eta,\zeta)\,(d\eta^2 - d\zeta^2) -
\beta^2(\eta,\zeta)\,(dx^2 + dy^2).
\ee
In practice we use the co-ordinate transformation to compute $A$ and
$B$ from \eq{metric}, as well as $\alpha$ and $\beta$, as we evolve
the system numerically.  The equations of motion will be given in
\cite{EastherET1999a}, and the numerical techniques used to solve them
are similar to those we used in \cite{ParryET1998a}. We focus on
initial configurations where a single mode (when viewed in the
perturbative context) is excited.  We choose $\alpha(0,\zeta) =
\beta(0,\zeta) = 1$, $\phi(0,\zeta)=\phi_0$ and $\alphaZ(0,\zeta)
=\betaZ(0,\zeta) = \PZ(0,\zeta)=0$. The constraints are solved if
\be
\alphaT(0,\zeta) = \frac{\kappa^2}{2C}\left( \frac{\PT^2}{2} +
V(\phi_0) \right) - \frac{C}{2}, \quad 
\betaT(0,\zeta) =
\sqrt{\frac{\kappa^2}{3} \left( \frac{\langle\PT^2\rangle}{2} +
V(\phi_0)\right)}= C. 
\ee
Our choice of $C$ ensures $\langle \alphaT(0,\zeta) \rangle =
\langle \betaT(0,\zeta) \rangle$, where $\langle \cdots \rangle$ is a 
spatial average. The actual inhomogeneity is injected through the
inflaton kinetic energy, 
\be \label{pert1}
\PT(0,\zeta) =  \Pi + \epsilon \sin{\left(\frac{2\pi k \zeta}{Z}\right)},
\ee
where $Z$ is the length of our ``box'', and $\Pi$ is the average
initial velocity. Since both $t$ and $z$ are transformed, slices of
constant $\eta$ in the conformal frame are not mapped directly to
slices of constant $t$ in the physical frame.  The initial density
perturbation is on the order of $\epsilon^2$.

\section*{Results}

\begin{figure}[tbp]
\begin{center}
\begin{tabular}{cc}
\psfrag{x}[t][]{$\zeta$}
\psfrag{y}[][]{$\eta$}
\psfrag{z}[][]{$$}
\includegraphics[scale=.65]{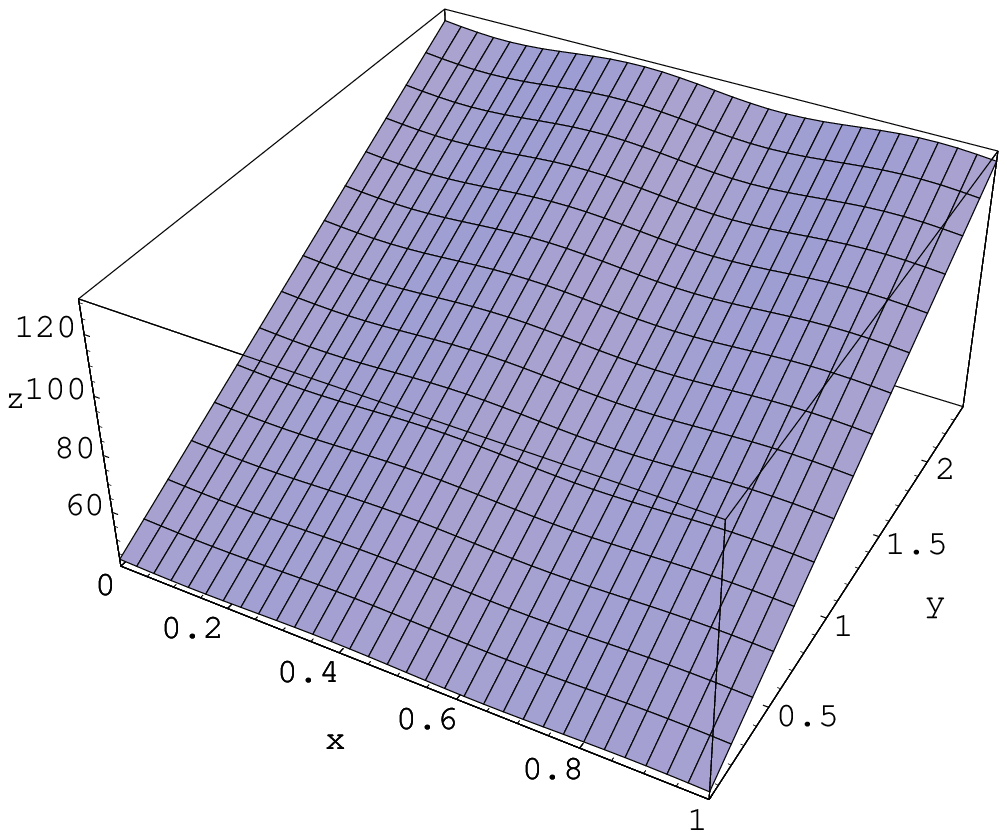} & 
\psfrag{x}[t][]{$\zeta$}
\psfrag{y}[b][]{$\eta$}
\psfrag{z}[][]{$$}
\includegraphics[scale=.65]{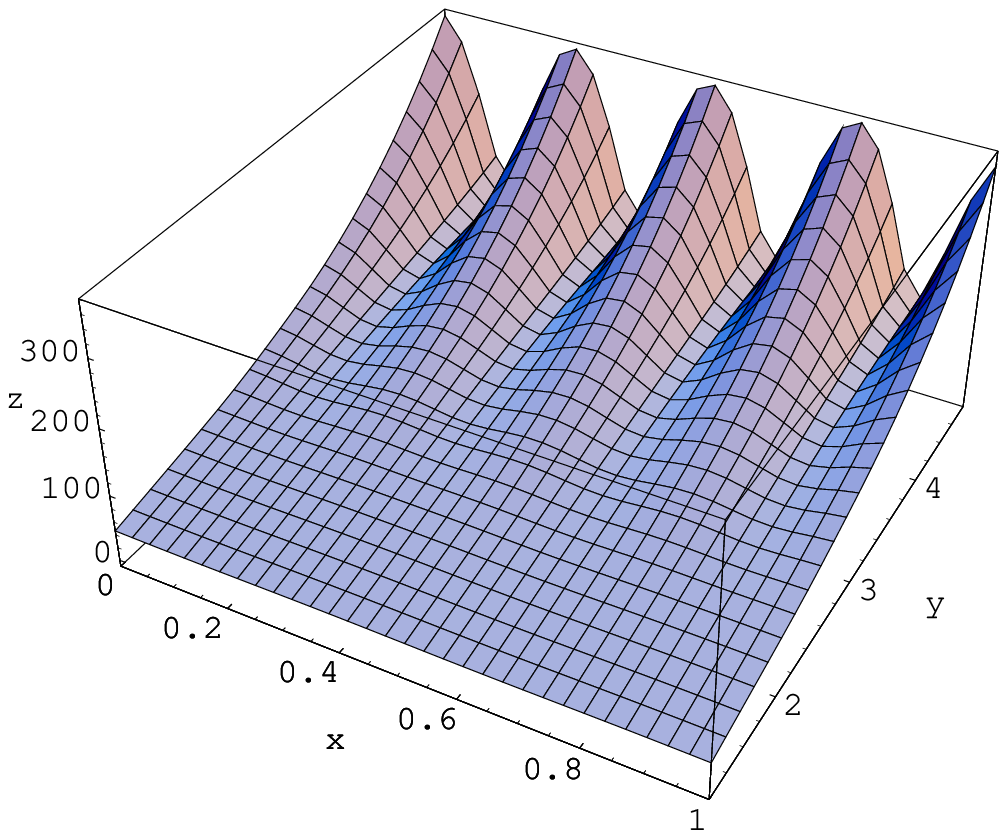}
\end{tabular}
\end{center}
\caption[]{The evolution of $A$ (the $g_{zz}$ component of \eq{metric})
is plotted as a function of $\eta$ and $\zeta$. The left panel shows
the evolution of a mode with $k$ slightly too large to be in
resonance, while the right hand case undergoes resonance, and
significant inhomogeneity is generated.  The units are arbitrary. The
initial perturbation is $\epsilon^2=10^{-6}$, and the simulations
begins at the end of inflation.}
\end{figure}

When we examined reheating after $m^2\phi^2$ we confirmed that there
is no significant resonant amplification, and that the perturbative
analysis is valid \cite{ParryET1998a}. However, after $\lambda\phi^4$
inflation there is a single, narrow resonance band, and the scaling
properties of the solution ensure that modes which are initially in
this band remain there indefinitely.

\begin{figure}[tbp]
\begin{center}
\begin{tabular}{cc}
\psfrag{x}[t][]{$\eta$}
\psfrag{y}[][]{$$}
\includegraphics[scale=.65]{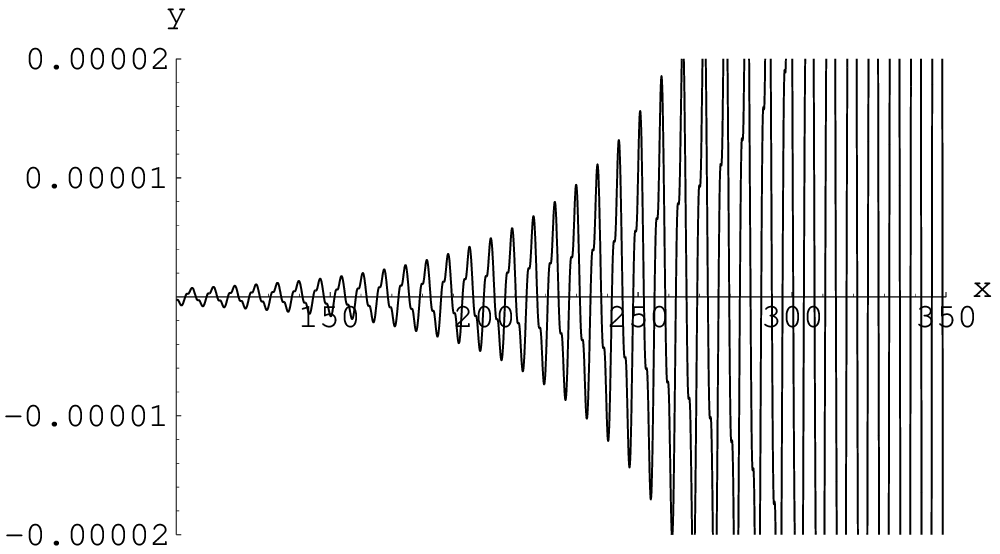} & 
\psfrag{x}[t][]{$\eta$}
\psfrag{y}[b][]{$$}
\includegraphics[scale=.65]{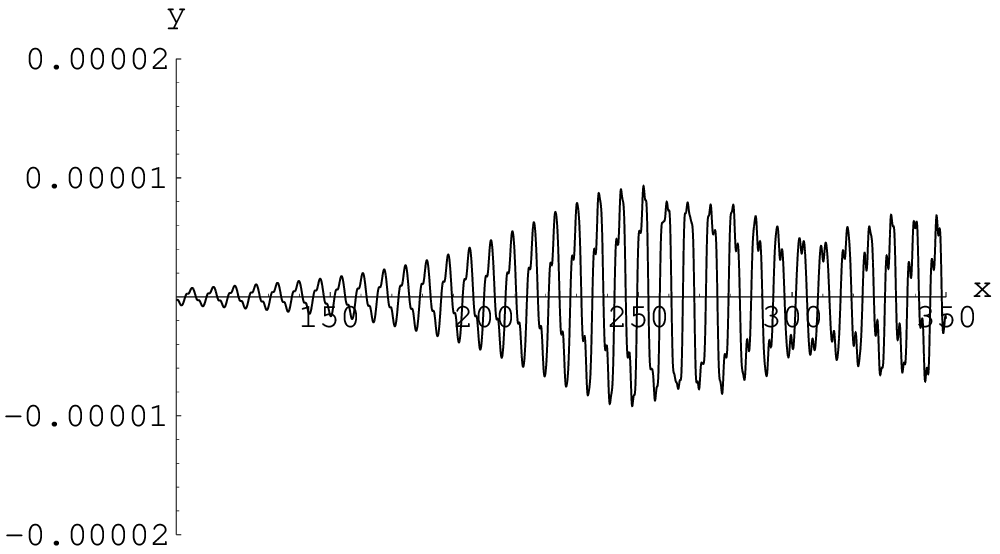}
\end{tabular}
\end{center}
\caption[]{The Fourier mode of $\Phi$, the metric perturbation, 
 for a resonant mode is shown: the left panel gives the
perturbative result, while the right panel shows the evolution of the
mode derived from the full nonlinear analysis.}
\end{figure}

Fig.~1 shows the evolution of the metric component $A$ for two
different perturbations, one inside the resonance band and one outside
it.  A significant degree of inhomogeneity is generated in the metric
by the resonance in the fields.  While a simple analytical treatment
suggests that resonance lasts indefinitely, the backreaction from the
nonlinear field evolution eventually halts the resonant amplification.
This feature is apparent in our simulation, and is illustrated in
Fig.~2, where we plot the metric perturbation, $\Phi$
\cite{MukhanovET1992b}. The termination of resonance coincides roughly
with the comparatively sudden growth of the metric perturbation seen
qualitatively in the right panel of Fig.~1.

If we solve for an inhomogeneous field, $\phi$, while averaging over
the spacetime background (corresponding to numerical treatments which
include the nonlinear field equations, but assume that the metric is
unperturbed) the subsequent evolution of $\Phi$ differs from that
which we obtain in the presence of the gravitational
backreaction. Comparing the two results will allow us isolate the
contribution from the inhomogeneous parts of the metric to the overall
evolution. We are also extending this work to more general initial
perturbations than \eq{pert1}, so we can include interactions between
different modes and study non-perturbative effects such as the
formation of primordial black holes after inflation
\cite{EastherET1999a}.

\section*{Acknowledgments}

We thank Robert Brandenberger and Fabio Finelli for useful
discussions.  Computational work in support of this research was
performed at the Theoretical Physics Computing Facility at Brown
University.  RE is supported by DOE contract DE-FG0291ER40688, Task A.

 \end{document}